\documentclass[12pt,preprint]{aastex}

\shorttitle{Optical Monitoring of OJ~287: a 40-Day Period?}
\shortauthors{Wu et al.}

\received{}
\accepted{}

\begin{document}

\title{Optical Monitoring of BL Lacertae Object OJ~287:\\ a 40-Day Period?}

\author{Jianghua Wu, Xu Zhou}
 \affil{National Astronomical Observatories, Chinese Academy of
        Sciences, 20A Datun Road, Beijing 100012, China}
 \email{jhwu@bao.ac.cn, zhouxu@bac.pku.edu.cn}

\author{Xuebing Wu, Fukun Liu}
 \affil{Department of Astronomy, Peking University, Beijing 100871, China}

\and

\author{Bo Peng, Jun Ma, Zhenyu Wu, Zhaoji Jiang, Jiansheng Chen}
 \affil{National Astronomical Observatories, Chinese Academy of
        Sciences, 20A Datun Road, Beijing 100012, China}

\begin{abstract}
We present the results of our optical monitoring of the BL Lacertae object
OJ~287 during the first half of 2005. The source did not show large-amplitude
variations during this period and was in a relatively quiescent state. A
possible period of 40 days was derived from its light curves in three BATC
wavebands. A bluer-when-brighter chromatism was discovered, which is
different from the extremely stable color during the outburst in 1994--96.
The different color behaviors imply different variation mechanisms in
the two states. We then re-visited the optical data on OJ~287 from the
OJ-94 project and found as well a probable period of 40 days in its optical
variability during the late-1994 outburst. The results suggest
that two components contribute to the variability of OJ~287 during its
outburst state. The first component is the normal {\sl blazar} variation. 
This component has an amplitude similar to that of the quiescent state and
also may share a similar periodicity. The second component can be taken as
a `low-frequency modulation' to
the first component. It may be induced by the interaction of the assumed
binary black holes in the center of this object. The 40-day period may be
related to the helical structure of the magnetic field at the base of the
jet, or to the orbital motion close to the central primary black hole.
\end{abstract}


\section{INTRODUCTION}

Blazars represent a peculiar subclass of active galactic nuclei (AGNs).
The most prominent property of blazars is their strong and rapid
variability, which is believed to originate from a relativistic jet that
is pointed basically towards an observer. Another characteristic of blazars
is their high polarization, with the degree and position angle also being
highly variable. Superluminal motions have been observed in a significant
fraction of these radio-loud flat-spectrum sources. Blazars can be
classified into flat-spectrum radio quasars and BL Lac objects, depending
on whether or not they show strong emission lines in their optical spectra.

Ever since the discovery of blazars and their highly variable brightness,
the efforts have never stopped in searching for a periodicity in their
variability. The reason for doing that is that the periodicity can put
strong constraints on the emission and variation mechanisms. Although most
attempts failed to find any periodicity, some positive results have been
claimed for several objects (e.g., OJ~287, \citealt{sillan88}; 3C~345,
\citealt{webb88}; 3C~120, \citealt{webb90}; S5~0716+714, \citealt{quir91};
ON~231, \citealt{liu95}; Mrk~421, \citealt{liu97};
PKS~0735+178, \citealt {fan97}; BL Lac, \citealt{fan98}; Mrk~501,
\citealt{haya98}; AO~0235+164, \citealt{raiteri01}). However, except for
the case of OJ~287, none of the claimed periods have been seen repeatedly,
and they appear to be only `transient periods'.

The BL Lac object OJ~287 is one of the best observed blazars. It is also
the only blazar that shows convincing evidence for periodic variations. By
good luck and its suitable location on the sky (very close to the ecliptic
where most asteroid and comet searches are made), its optical photometric
measurements can be dated back to more than 100 years ago. The most prominent
feature in its historical light curves is the cyclic outbursts with an
interval of about 12 years, based on which \citet{sillan88} proposed a
binary black hole (BBH) model for this object and predicted that a new
outburst would occur in late-1994. In order to verify the predicted outburst,
an international project was organized to monitor OJ~287 in multi-wavebands.
This is the OJ-94 project, which covered the time range from fall 1993 to
the beginning of 1997. The predicted outburst was observed with one peak
at 1994.8 and another at 1996.0 \citep{sillan96a,sillan96b}. This result
confirms the 12-year periodicity in the optical variability of OJ~287.

The OJ-94 project found a double-peaked structure and a quite stable color
for the major outburst \citep{sillan96a,sillan96b}. Therefore, the ``old"
BBH model by \citet{sillan88} had to be modified, since it cannot explain
the double-peaked structure. New models include the ``old" hit and penetration
model by \citet{lehto96}, the precessing disk/jet model by \citet{katz97}, and
the beaming model by \citet{villata98}. These new models also require a BBH
system in the center of OJ~287, and can well explain the 12-year period,
double-peaked structure of the outburst, and/or stable color \citep[see a
review by][]{sillan96b}. However, radio and polarization observations
\citep{valtaoja00,pursimo00} show that the first peak in late-1994 is a
thermal flare lacking a radio counterpart, while the second peak in 1995--96
is apparently a flare dominated by synchrotron radiation and is
accompanied by a radio outburst. Two previous outbursts in 1971--73 and
1983--84 also have this property. These results can not be explained by the
three new models mentioned above. By incorporating the radio and polarization
results, \citet{valtaoja00} suggested a new hit and penetration model, in
which the secondary black hole hits and
penetrates the accretion disk of the primary during the pericenter passage,
causing a thermal flare visible only in the optical regime. At the same
time, the pericenter passage enhances accretion into the primary black hole,
leading to increased jet flow and formation of shocks down the jet. These
become visible as simultaneous radio-optical synchrotron flares and are
identified with the second optical peaks. Later, \citet{liu02} derived
detailed parameters of the BBH system and estimated the mass of the primary
black hole as $4\times10^8\,M_{\odot}$.

Alternatively, in order to explain the lack of a simultaneous radio flare
in the late-1994 outburst, \citet{marscher98} proposed that the
late-1994 outburst comes from the base of the jet, near the central engine,
while the simultaneous radio-optical flare in 1995--1996 occur in the radio
core region, about a parsec down the jet. Since the base of the jet must be
utterly opaque to radio emission, the first flare is not observed in radio
regimes. He also mentioned that a ``duty cycle" of winding up of the magnetic
field at the base of the jet would also result in major quasi-periodic
injection of enhanced flow into the jet \citep{ouyed97} and hence the
observed periodic outbursts.

In order to re-verify the 12-year period and to evaluate the various models
on OJ~287, more intensive monitoring should be carried out, not only during
the outburst phases, but also in its quiescent states. We monitored
OJ~287 in the first half of 2005, about 1.5 year before the predicted next
outburst \citep{valtonen97,kidger00,valtaoja00,liu02}. The aims are
to record the variability in its quiescent state,
to prepare a comparison to the variability in its outburst
state, and to give more constraints to its physical model.
Here we present our monitoring results and compare them with those of the
1994--96 outburst observed in the OJ-94 project. Section 2 describes our
observations and data reduction procedures. The results are presented in
\S3, and \S4 describes our re-analyses of the OJ-94 data. The
physical processes responsible for the variability are discussed in \S5,
and a summary is given in \S6.

\section{OBSERVATIONS AND DATA REDUCTION}

Our optical monitoring of OJ~287 was carried out on a 60/90 cm Schmidt
telescope located at the Xinglong Station of the National Astronomical
Observatories of China (NAOC). A Ford Aerospace $2048\times2048$ CCD camera
is mounted at its main focus. The CCD has a pixel size of $15\,\micron$ and
a field of view of $58\arcmin\times58\arcmin$, resulting in a resolution of
$1\farcs7\,\rm{pixel}^{-1}$. The telescope is equipped with a 15 color
intermediate-band photometric system, covering a wavelength range from
3000 to 10,000 \AA. The telescope and the photometric system are mainly
used to carry out the Beijing-Arizona-Taiwan-Connecticut (BATC) survey
\citep{zhou05}.

The monitoring covered the time from 2005 January 29 to April 28, or from
JD~2,453,400 to 2,453,489. As a result of weather conditions and
observations of other targets, there are actually 27 nights' data in total.
We used the BATC $e$, $i$, and $m$ filters. Their central wavelengths
are 4885, 6685, and 8013 \AA, respectively\footnote{The BATC $i$ and $m$
bands are close to the Cousins $R$ and $I$ bands, respectively. One can
transfer the BATC $i$ magnitudes to the Cousins $R$ magnitudes with the
perfect linear relation $R=i+0.1$ for normal stars \citep{zhou03}.}. In most
nights, we made
photometric measurements in only one cycle of the BATC $e$, $i$, and $m$
bands, while in a small fraction of nights, more cycles of exposures were
made. The exposure times are mostly 240 s in the BATC $e$ and $m$ bands and
150 s in the $i$ band. The observational log and parameters are presented
in Tables~1--3.

The procedures of data reduction include positional calibration, bias
subtraction, flat-fielding, extraction of instrumental aperture magnitude,
and flux calibration.
The average FWHM of
stellar images was about 4\farcs5 during our monitoring. So the
radii of the aperture and the sky annulus were adopted as 5, 7, and 10
pixels (or 8\farcs5, 12\arcsec, and 18\arcsec) respectively during the
extraction. We used the comparison stars 4, 10, and 11 in \citet{fio96} for
the flux calibration of OJ~287. Their BATC $e$, $i$, and $m$ magnitudes are
obtained by observing them and three BATC standard stars, HD~19445, HD~84937,
and BD+17d4708, on a photometric night, and are listed in Table~\ref{T4}.
Then, by comparing the instrumental magnitudes of the three comparison stars
with their standard BATC magnitudes, the instrumental magnitudes of OJ~287
were calibrated into the BATC $e$, $i$, and $m$ magnitudes, and the light
curves in the three BATC bands were obtained.

\section{RESULTS}

\subsection{Light Curves}

\begin{figure}
\plotone{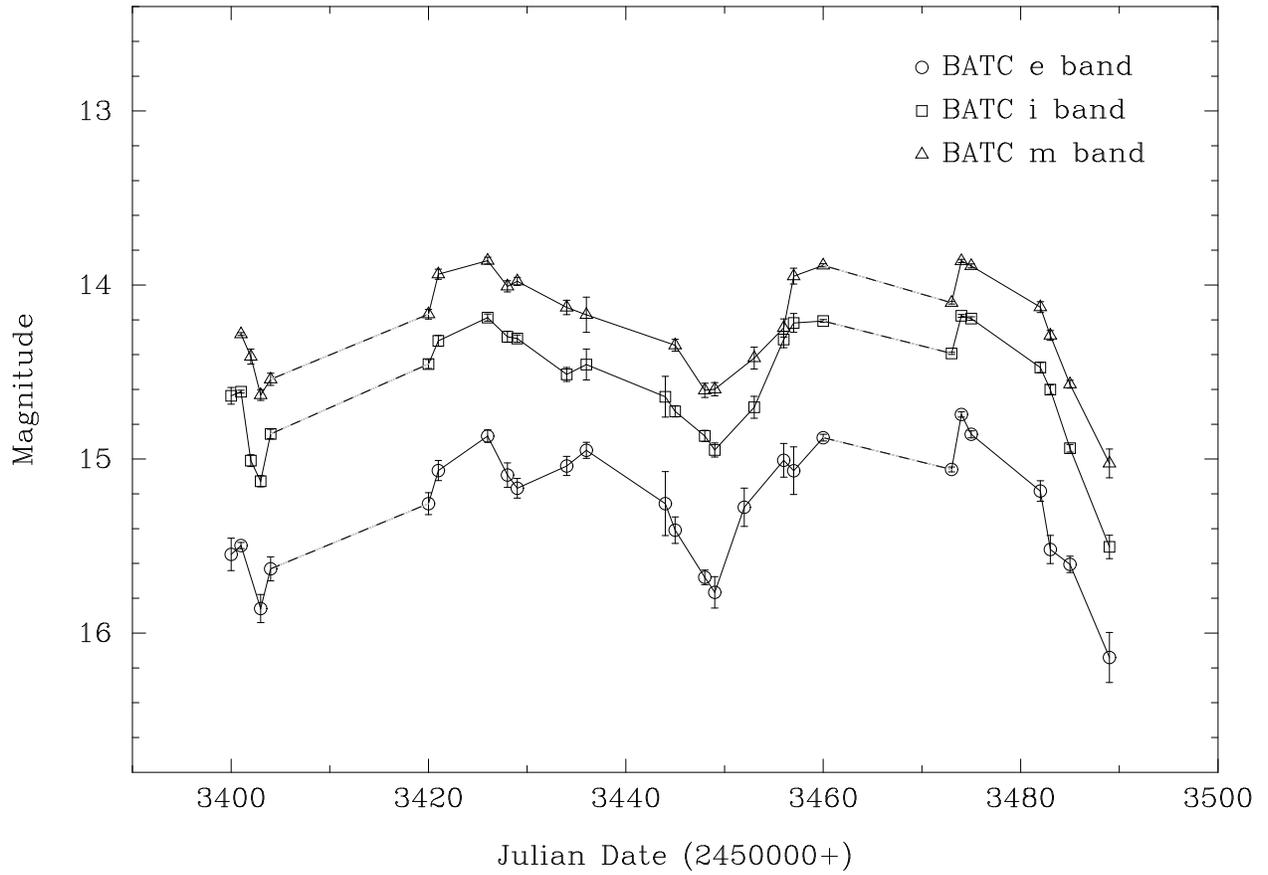}
\caption{The light curves in the BATC $e$, $i$, and $m$ bands. Only
nightly-mean magnitudes are plotted.}
\label{F1}
\end{figure}

The light curves in the three BATC bands are displayed in Figure~\ref{F1}.
Here we plot only the nightly-mean magnitudes, since the amplitudes of
variations during all individual nights (if there are multi-cycles of
exposures, see \S2) are mostly less than 0.2 mag. The variations in the
three BATC bands are basically consistent with each other. The overall
amplitude was about 1.3 mag during the whole monitoring period, and the
object was in a relatively quiescent state, as expected. Two cycles of
variations show up in the light curves: The object varied from a minimum on
JD~2,453,403 to a maximum on JD~2,453,426, and then went back to a new
minimum on JD~2,453,449. With a sharp turnover, the object brightened
again and reached a second maximum around JD~2,453,474. After that, the object
dropped its brightness again to a third minimum on JD~2,453,489 (see also
Tables~1--3). The time intervals are 46 and 40 days between the successive
minima, and 48 days between the two maxima. The average is 44 days, which
could be taken as the period of the variations.

One may argue that the minimum on JD~2,453,489 may not represent the actual
end of the second cycle. But one can see from Figure~\ref{F1} that
the end parts of the light curves have very steep slopes, which implies
that the object might get still fainter in magnitude, but will not spend too
much in time to reach the apparent end of the second ``cycle". In other words,
it appears probable that JD 2,453,489 is close to, if not actually at, the
end of a second cycle.

\subsection{Period of Variability}

Visual inspection of the light curves in Figure~\ref{F1} indicates a period
of 44 days in the variations of OJ~287. In order to quantitatively derive
the period, we performed a structure function (SF) analysis on the light
curves. SF is frequently used to search for the typical timescales and
periods in the variability \citep{simon85}. A characteristic timescale in a
light curve, defined as the time interval between a maximum and an adjacent
minimum or vice versa, is indicated by a maximum of the SF, whereas a period
in the light curve causes a minimum of the SF \citep{smith93,heidt96}. SF is
usually calculated twice by using an interpolation algorithm, first starting
from the beginning of the time series and proceeding forwards, and secondly
starting from the end and proceeding backwards. This may result in two
slightly different SF curves but will provide a rough assessment of the
errors caused by the interpolation process.

\begin{figure}
\plotone{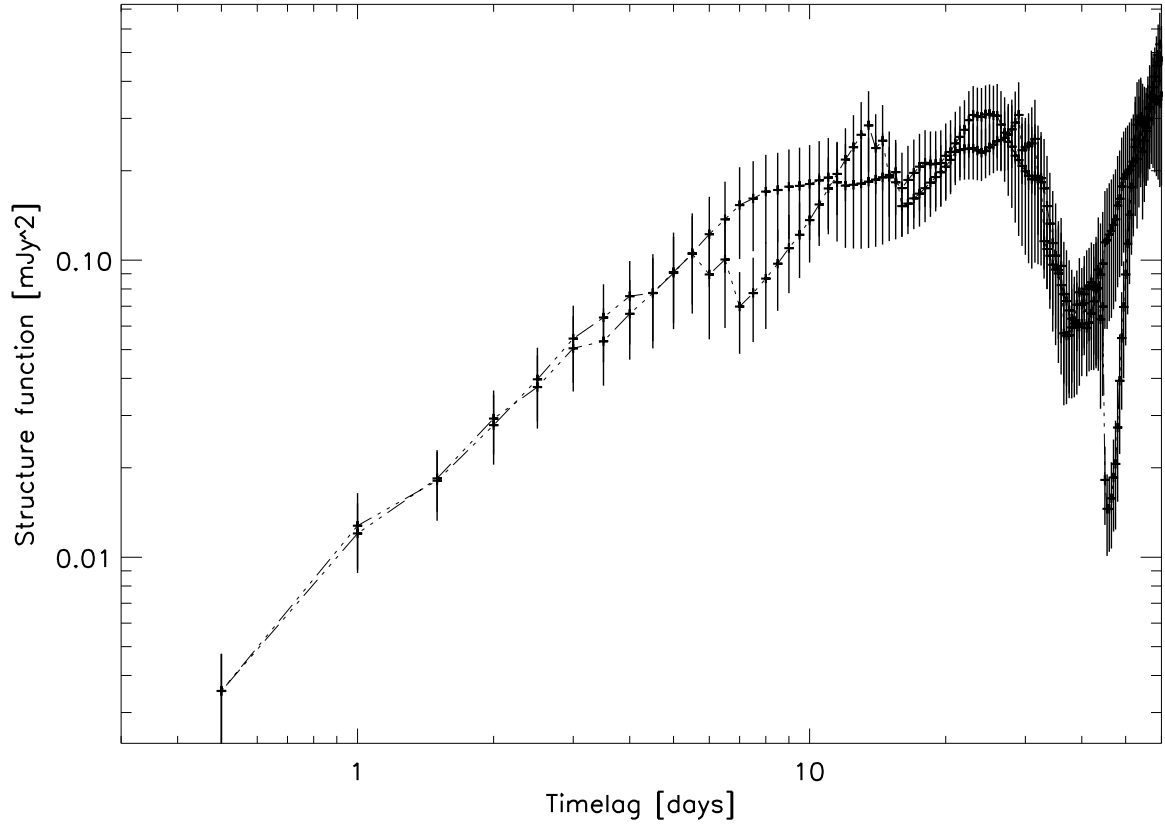}
\caption{Structure function of the light curve in the BATC $i$ band. The
minima at 34 and 44 days indicate the periods of the variation. See text for
details.}
\label{F2}
\end{figure}

Figure~\ref{F2} shows the SF of the light curve in the BATC $i$ band. There
is a deep minimum at about 44 days, which confirms the 44-day period
estimated with the above visual inspection. Besides the minimum at 44 days,
there is a secondary minimum around 34 days
on the SF curve. It should reflect the time interval between JD~2,453,426 and
JD~2,453,460, the two consecutive maxima on the light curves (while the
44-day period mainly reflects the time intervals between consecutive minima).
SF curves in the other two BATC bands also show both these `periods'.
In principle, the time intervals between any two consecutive in-phase points
in a periodic light curve should be equal to each other and equal to the
period. Here the difference between the two `periods' may be the result
of the unevenly sampled data (for example, the actual second maximum may be
between JD~2,453,460 and JD~2,453,473 where we have no observations) and
the relatively short time coverage of our monitoring. The two periods are
expected to converge in a longer and more evenly sampled monitoring program.
So here we take the mean of them, $\sim40$ days, as the actual period in the
variations.

Although both visual inspection and SF analysis indicate a period of around
40 days, it must be noted that our observations cover only two cycles of that
apparent period. Moreover, in the
light curves there are two gaps (JD~2,453,404--420 and JD~2,453,460--473),
which may include additional maxima and/or minima. So this 40-day period needs
to be confirmed with future observations, but it does lead us to search for
this possible period in the earlier data. Indeed, much stronger evidence for
the 40-day period was found, which will be reported in \S4.

Besides the 40-day period reported here and the prominent 12-year period,
\citet{fan02} have reported a period of 5.53 years for the optical
variability of OJ~287. On shorter timescales, \citet{efimov02} observed an
apparent period of 36.56 days for the rotation of the position angle of the
optical polarization. Small fluctuations in intensity with periods of 10--20
min have also been claimed \citep{carrasco85,diego90} but have been
(at best) of a transient nature. Our 40-day period is somewhat consistent
with the period reported by \citet{efimov02}, which will be discussed later.

\subsection{Spectral Behavior}

Spectral behavior involved in the variability of blazars can put strong
constraints on their variation mechanisms, as demonstrated by \citet{wu05}.
Optical spectral changes with brightness have been investigated for a number
of blazars \citep[e.g.,][]{carini92,ghisellini97,speziali98,romero00,
villata02,villata04,raiteri03,vagnetti03,wu05}. Most authors have reported
a bluer-when-brighter
chromatism when the objects show fast flares and an achromatic trend for
their long-term variability. OJ~287 was found to have an extremely
stable color during the 1994--96 outburst \citep{sillan96b}. Here we
investigate its spectral behavior in its relatively quiescent state.

\begin{figure}
\plotone{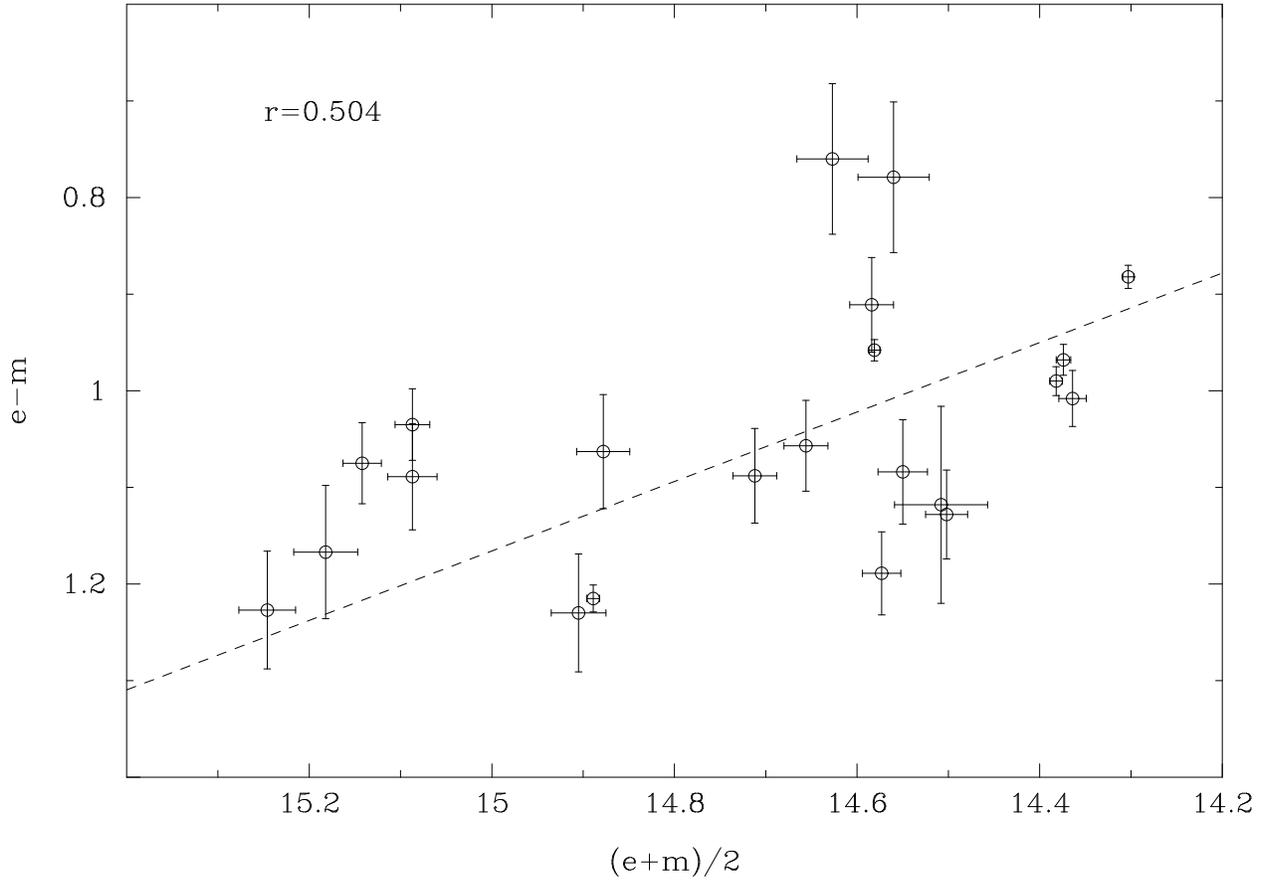}
\caption{Color index vs. brightness. The dashed line is the best fit to the
points. There is a clear bluer-when-brighter chromatism.}
\label{F3}
\end{figure}

Following \citet{raiteri03} and \citet{wu05}, we use color index to denote
spectral shape, and calculate the color as $e-m$ and brightness as
$(e+m)/2$ for the BATC intermediate-band photometric system. As in \S3.1
and \S3.2, the nightly-mean magnitudes were used to calculate the colors
and brightness. Figure~\ref{F3} displays the color--brightness dependence.
The dashed line is the best fit to the points, with the errors in both
coordinates been taken into account \citep{press92}. The Pearson
correlation coefficient is 0.504 and the significance level is 0.017. So
there is a significant correlation between the brightness and color index,
or in other words, there is a clear bluer-when-brighter chromatism. This is
consistent with the bluer-when-brighter trend found by \citet{vagnetti03},
but is different from the extremely stable color during the outburst in
1994--96 \citep{sillan96b}.

The different spectral behaviors between the quiescent and outburst states
may indicate different variation mechanisms. In fact, in the quiescent state,
the essentially simultaneous optical and radio small flares \citep[e.g.,][]
{pursimo00,valtaoja00} and the bluer-when-brighter chromatism found in this
work support the hypothesis that shocks propagating along the relativistic
jet and interacting
with the hydrodynamically turbulent plasma and twisted magnetic field should
be responsible for the variations in the quiescent state \citep{wagner95,
marscher98}. On the other hand, it is very likely that the bulk increase in
brightness during the outburst state is the result of the impact of the
secondary black hole onto the primary accretion disk and the subsequent
enhanced accretion \citep{valtaoja00}, as mentioned in \S1.

\section{THE 1994--96 OUTBURST REVISITED}

After deriving a possible period of 40 days and obtaining the properties of
the variability of OJ~287 in its quiescent state, we then tried to search
for new proofs to the period and compare the properties to those in the
outburst state. The predicted next outburst is in 2006 \citep{valtonen97,
kidger00,valtaoja00,liu02}. So we at first look into the data of the outburst
in 1994--96. The data on the 1994--96 outburst are from the archive of
the OJ-94 project\footnote{http://www.astro.utu.fi/oj94/}. With a visual
inspection on the optical light curves of the outburst during
1994.7--1995.5 \citep[e.g., Fig.~1 in][]{sillan96b}, we found that there were
some small amplitude flares occurring at intervals of about 40 days overlaid
on the prominent outburst. This is in excellent agreement with the 40-day
period found in our monitoring program. We then analyzed the data in detail.

Our data analyses focused on the Johnson and Cousin $V$, $R$, and $I$ bands,
which are the most densely sampled wavebands in the OJ-94 project. We at
first analyzed the light curves from 1994.7 to 1995.5. In order to show the
small flares more clearly, we at first made a Fast Fourier Transform (FFT)
smoothing to the light curves. To have better smoothing results, the light
curves were truncated at both ends where the samplings are very low. The
smoothed light curves were then subtracted from the original ones, and the
`residual light curves (or variations)' were obtained. The procedure is
illustrated in Figure~\ref{F4}. The large panels display the original light
curves (pluses) and the smoothed ones (solid lines), while the small panels
show the residual light curves. Here we carried out 240-, 100-, and 140-point
FFT smoothings to the original $V$-, $R$-, and $I$-band light curves,
respectively. In all three small panels, the flares can be seen clearly
around JD~2,449,638, 2,449,670, 2,449,717, 2,449,752, 2,449,790, and
2,449,832. Except for the first flares, all other consecutive flares have
intervals of about 40 days. The mean variation amplitude of the flares is
about 1.0 mag, similar to that during the quiescent state (see \S3.1).

Also notable is that there seems to be some sub-flares between the major
flares mentioned above, i.e., at around JD~2,449,655, 2,449,700, 2,449,735,
2,449,775, and 2,449,817. They are weaker but somewhat broader at peaks
than the major flares. The time intervals between them are also about
40 days, but, of course, the intervals between them and the neighboring
major flares are about 20 days.

In order to derive the period quantitatively, the SFs and $z$-transformed
discrete correlation functions \citep[ZDCFs,][]{alex97} were calculated (in
auto-correlation mode for the ZDCFs) for the residual light curves. The
results are displayed in Figure~\ref{F5}. All three SF curves have a deep
minimum at about 40 days, and all three ZDCF curves show peaks at 40, 80,
and 120 days. Both indicate a period of 40 days. That is to say, the SF and
ZDCF analyses confirmed the 40-day period found with visual inspections.
This 40-day period is in excellent agreement with the 40-day period reported
in \S3.2.

We then investigated the spectral behaviors of the residual variations.
Figure~\ref{F6} displays the $(\Delta V-\Delta R)$ versus $\Delta V$ (left)
and $(\Delta V-\Delta I)$ versus $\Delta V$ (right) distributions. As in \S3,
we used the nightly-mean `residual magnitudes' to denote the `brightness' and
to calculate the `color'. There are strong bluer-when-brighter chromatisms
in both brightness--color diagrams. The dashed lines are the linear fits to
the points. The Pearson correlation coefficients are respectively 0.506 and
0.488, and the significance levels are $8.07\times10^{-11}$ and
$1.74\times 10^{-9}$, which indicate very strong correlations between the
color and brightness. These bluer-when-brighter chromatisms are again in
agreement with the color behavior of OJ~287 during its relatively quiescent
state.

After investigating the variations of the first peak in late-1994, we then
checked the data piece of the second peak in 1995--96 and those before and
after the two peaks. The second peak and the data after it do not show a
similar period, while the light curves before the first peak show some signs
of quasi-period oscillations（QPOs). Figure~\ref{F7} displays the $V$-band
light curve. Flares can be seen at JD~2,449,250, 2,449,293, 2,449,327, and
2,449,366, with intervals of about 40 days. After that, the light curve is
characterized by three strong sharp flares (at JD~2,449,366, 2,449,415, and
2,449,476--482) separated by three weaker but broader flares (at
JD~2,449,386--397, 2,449,440--459, and 2,449,501--514), a pattern very
similar to that in the late-1994 outburst. The intervals between the sharp
flares is 50--60 days. Because of the unevenly sampled data and the
apparently changing intervals, we do not perform a quantitative assessment
to this portion of the data, but the light curves may show QPOs.

\section{Physics of the 40-Day Period}

\begin{figure}
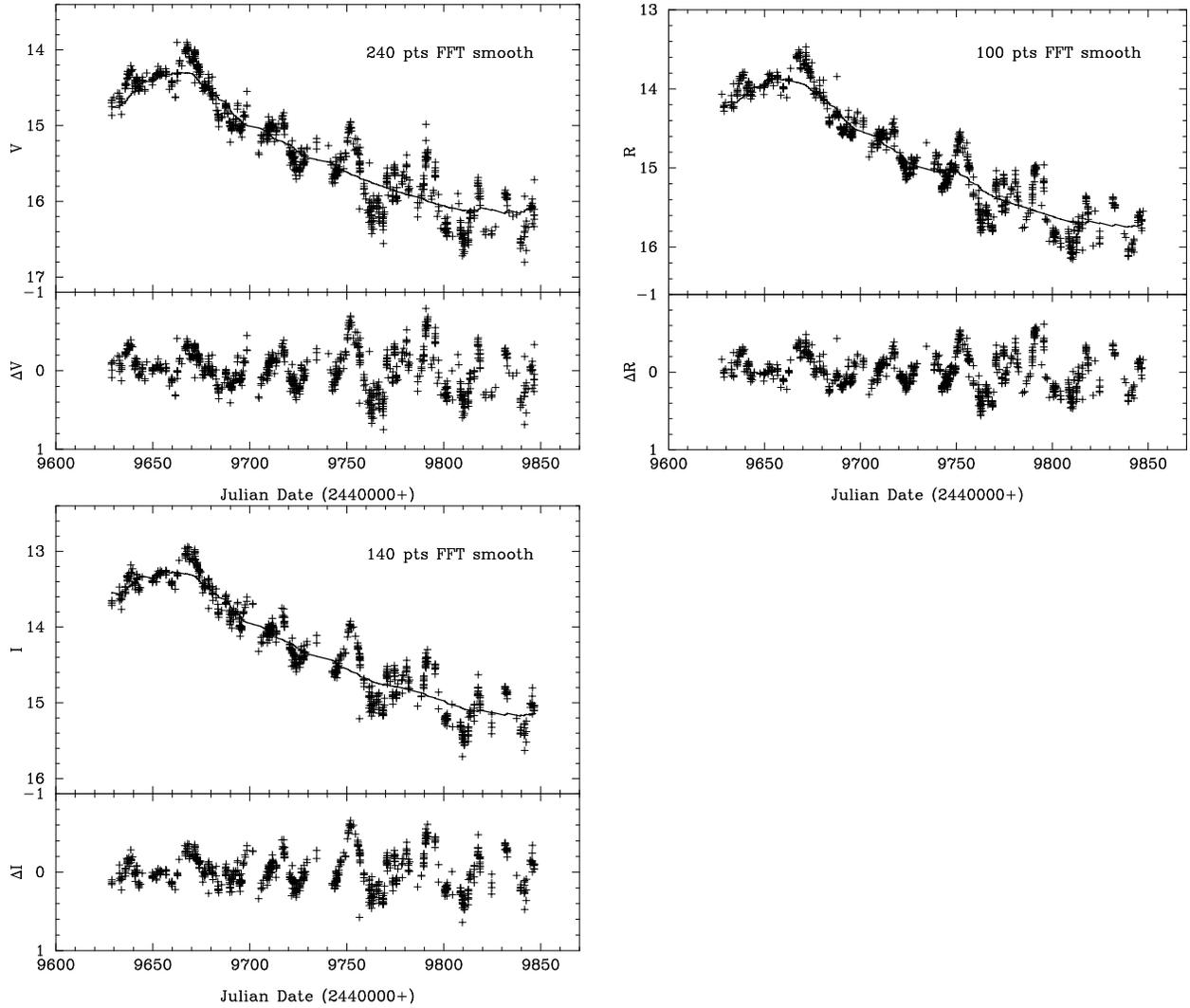

\includegraphics[height=7cm]{f4a.ps}
\includegraphics[height=7cm]{f4b.ps}
\includegraphics[height=7cm]{f4c.ps}
\caption{Original (larger panels) and residual (smaller panels) light curves of
the late-1994 outburst in the Cousins $V$, $R$, and $I$ bands, respectively.
The solid curves are respectively the 240-, 100-, and 140-point FFT smoothings
to the original light curves. The residuals are the differences between the
originals and the smoothings.}
\label{F4}
\end{figure}

\begin{figure}
\includegraphics[height=6cm]{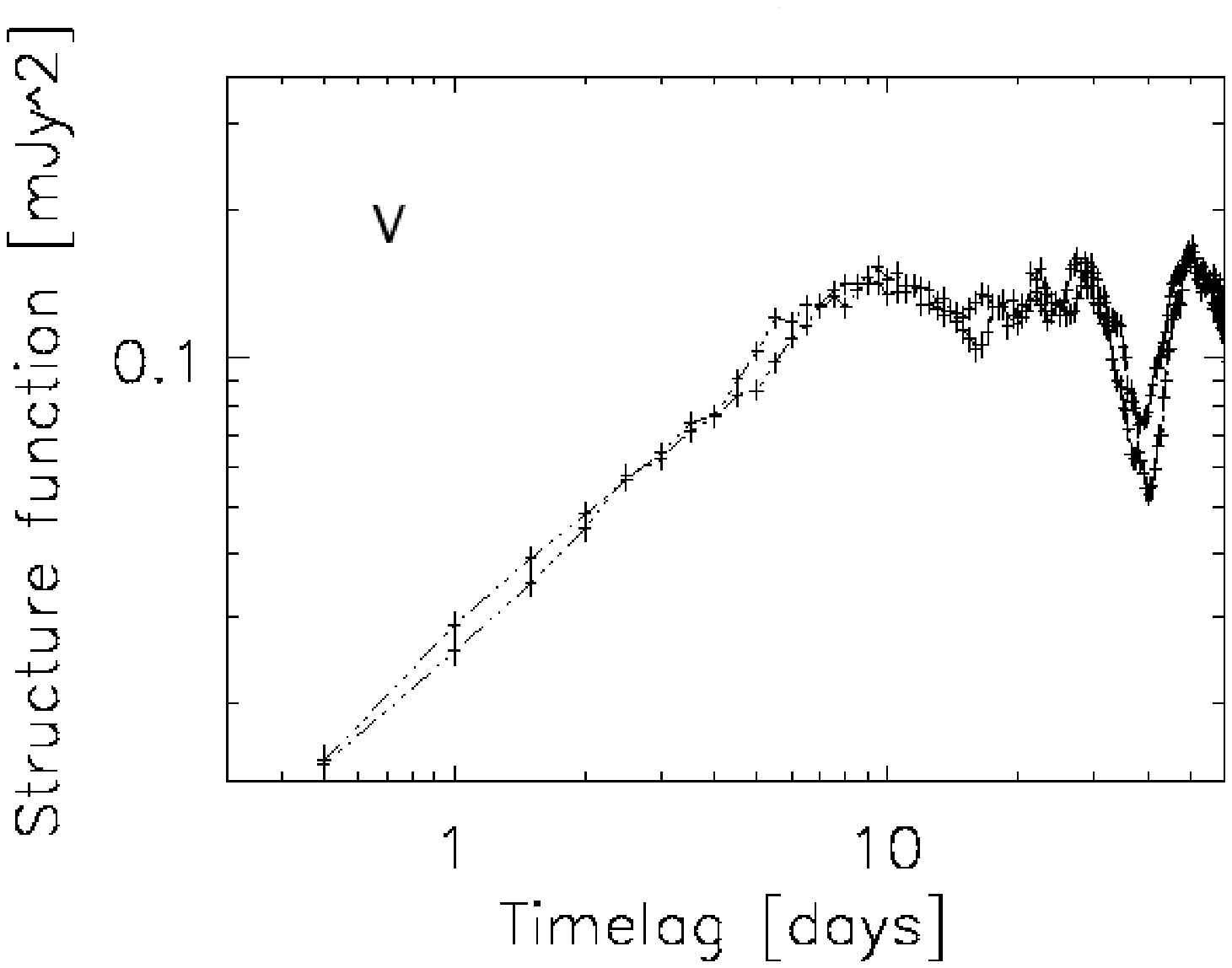}
\includegraphics[height=6cm]{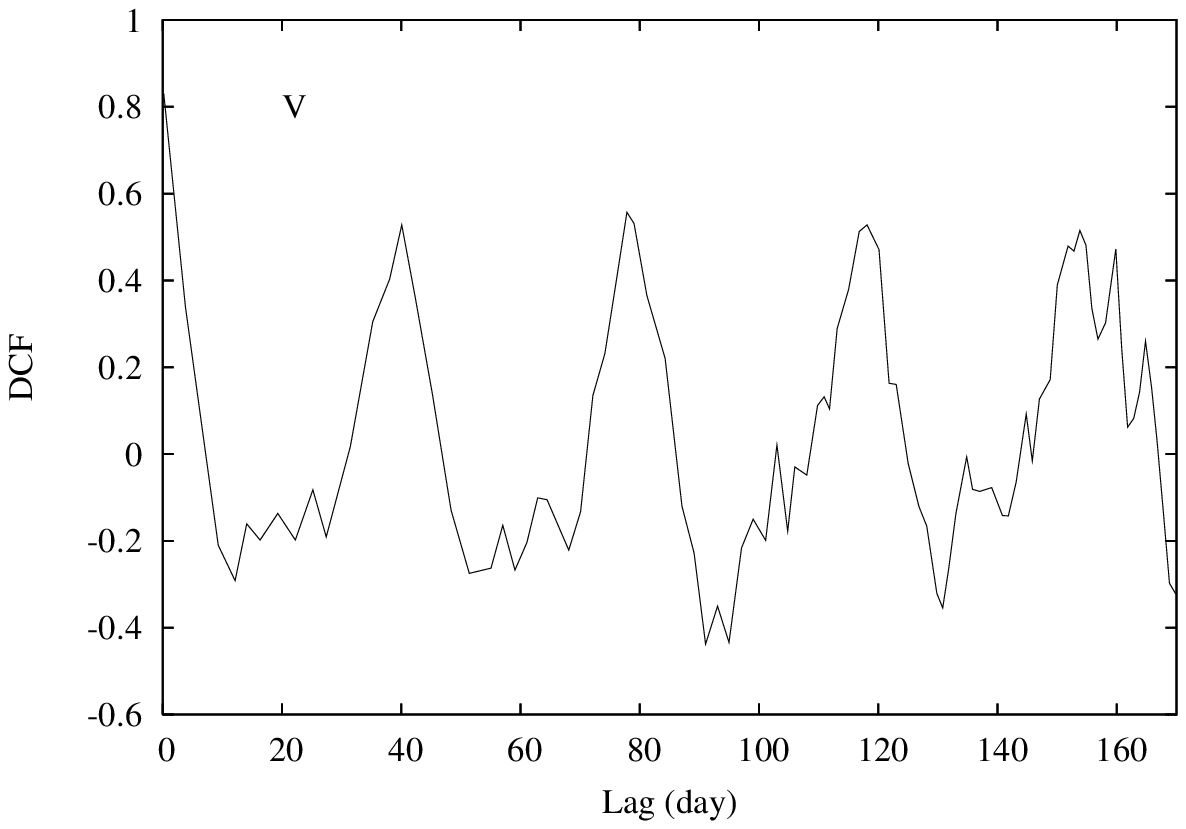}
\includegraphics[height=6cm]{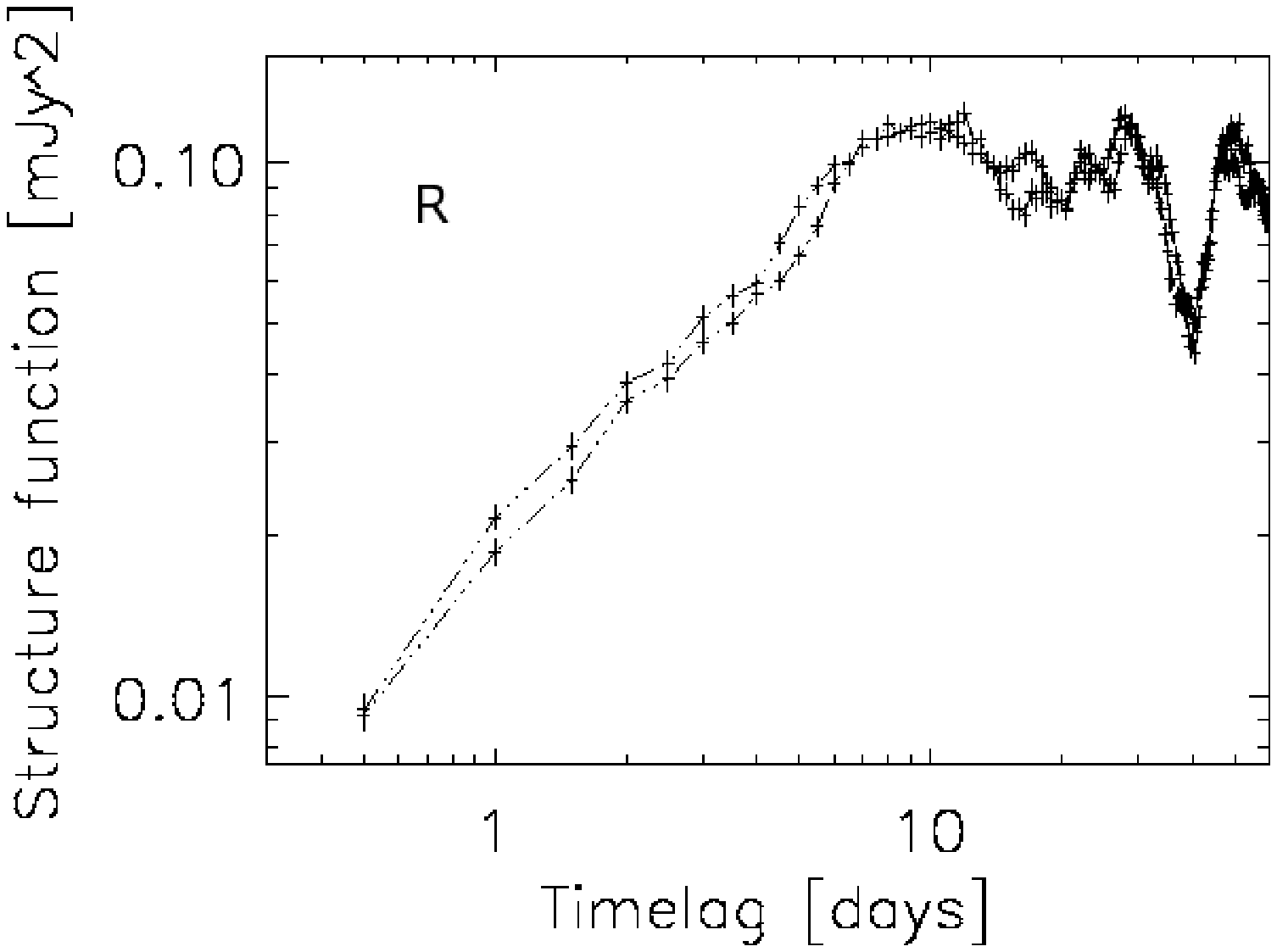}
\includegraphics[height=6cm]{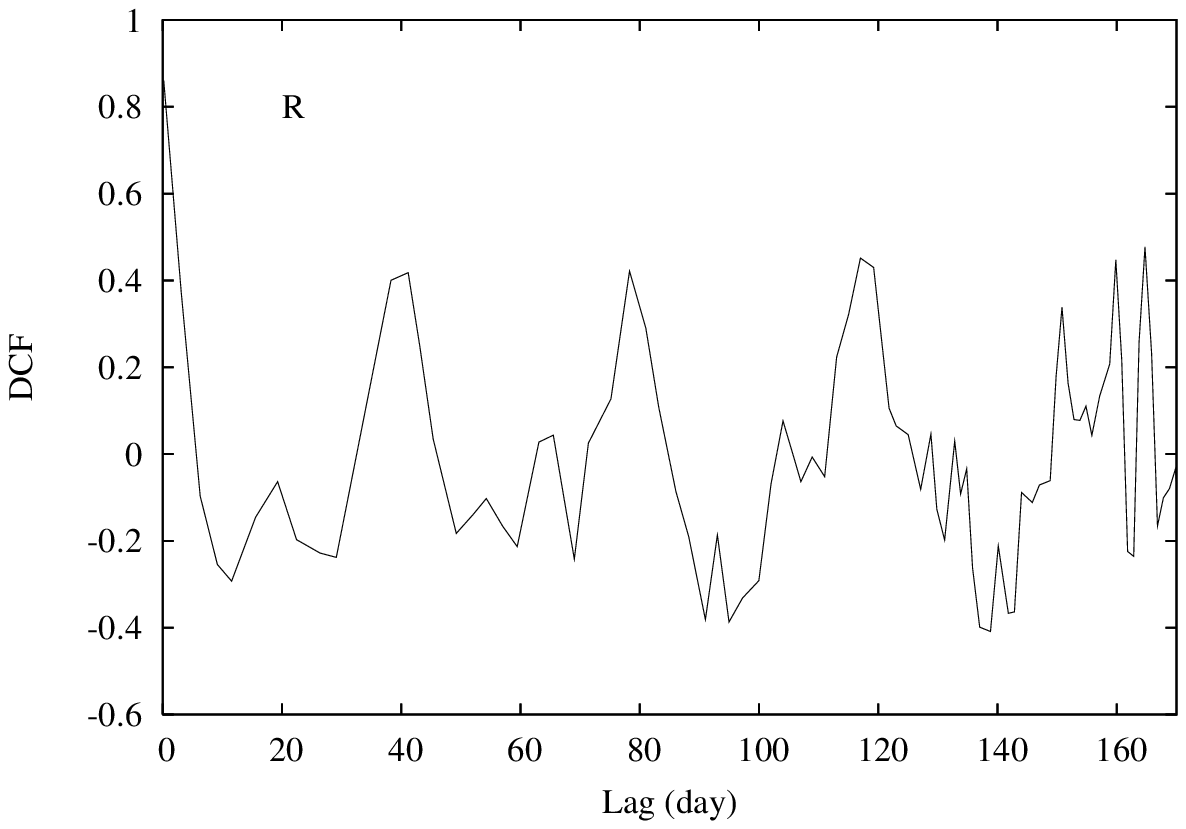}
\includegraphics[height=6cm]{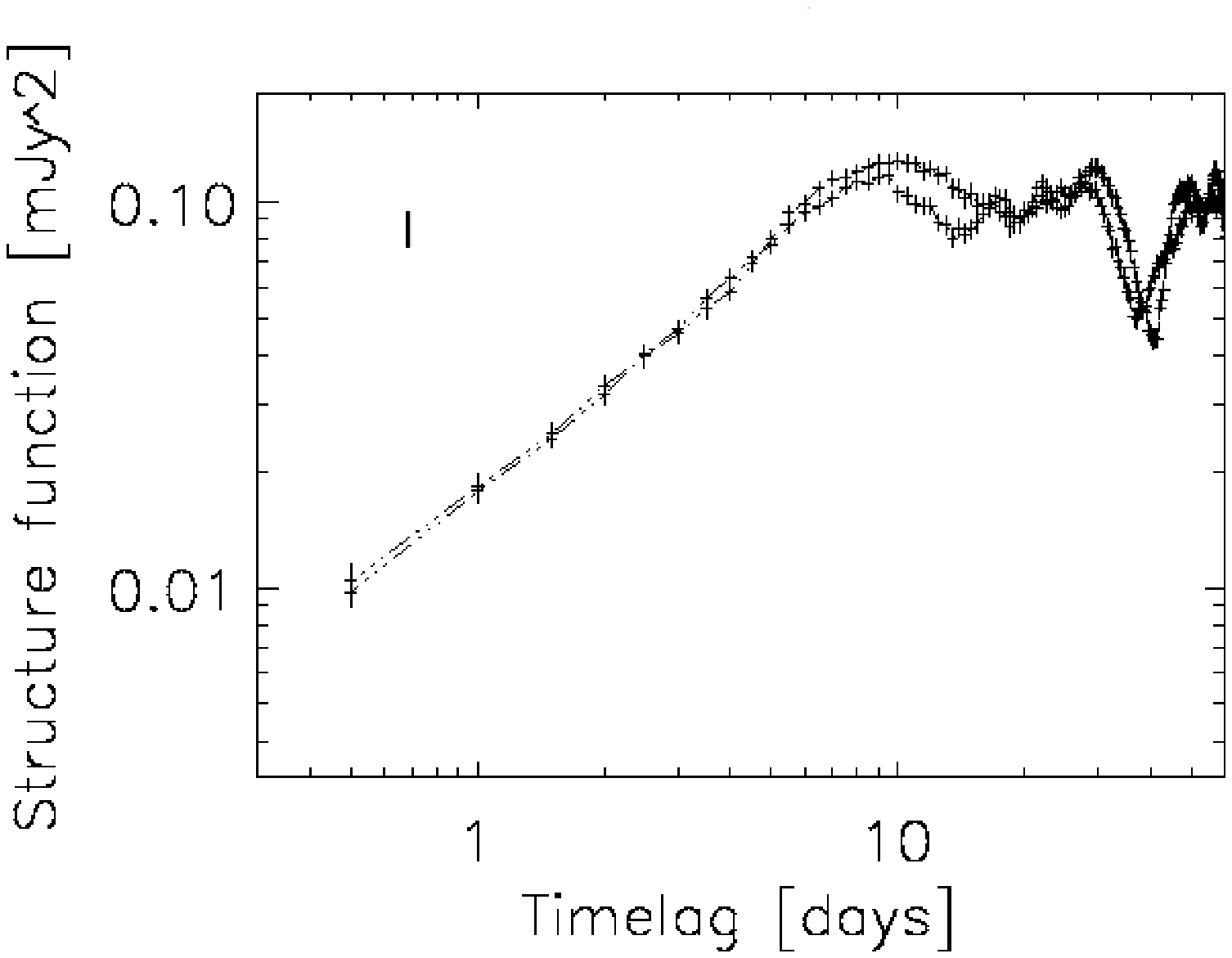}
\includegraphics[height=6cm]{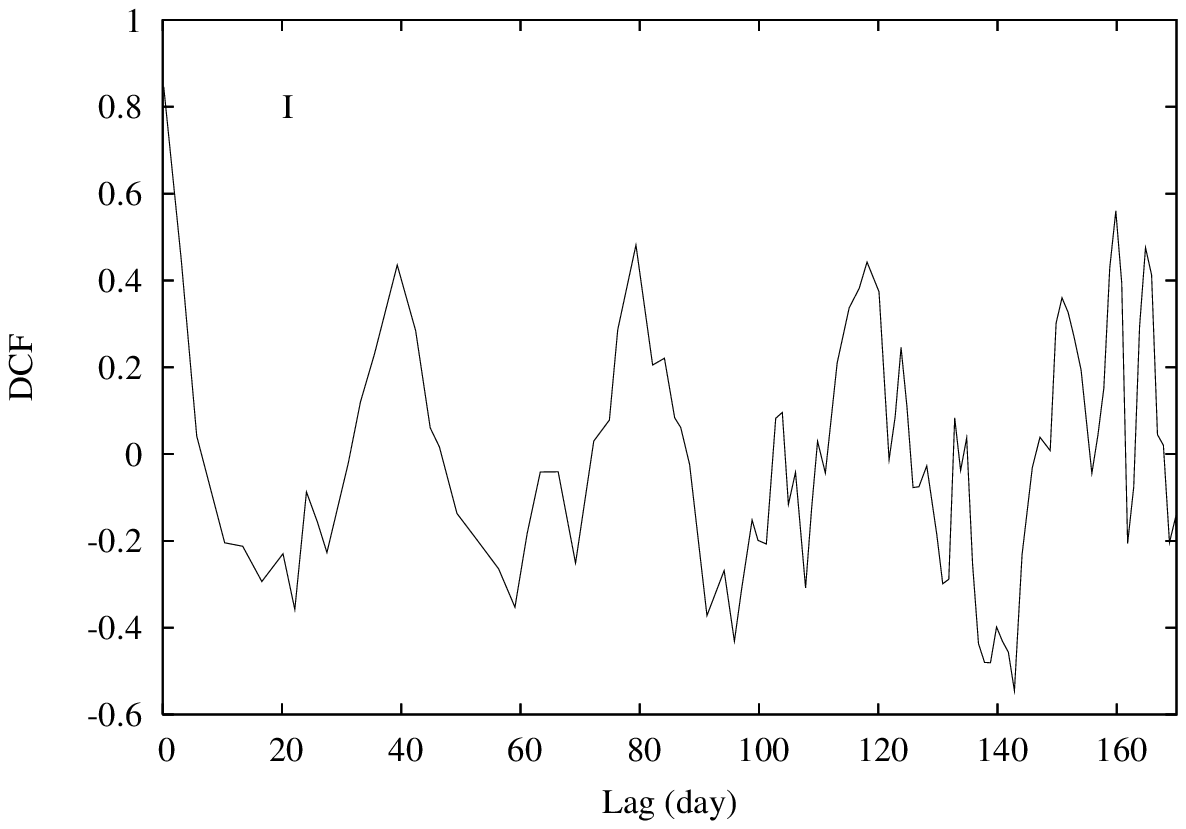}
\caption{SFs (left) and ZDCFs (right) of the residual light curves in
Fig.~\ref{F4}. There is a deep minimum around 40 days in all SFs, and the
ZDCFs show peaks around 40, 80, and 120 days, which both indicate a period
of about 40 days.}
\label{F5}
\end{figure}

\begin{figure}
\plottwo{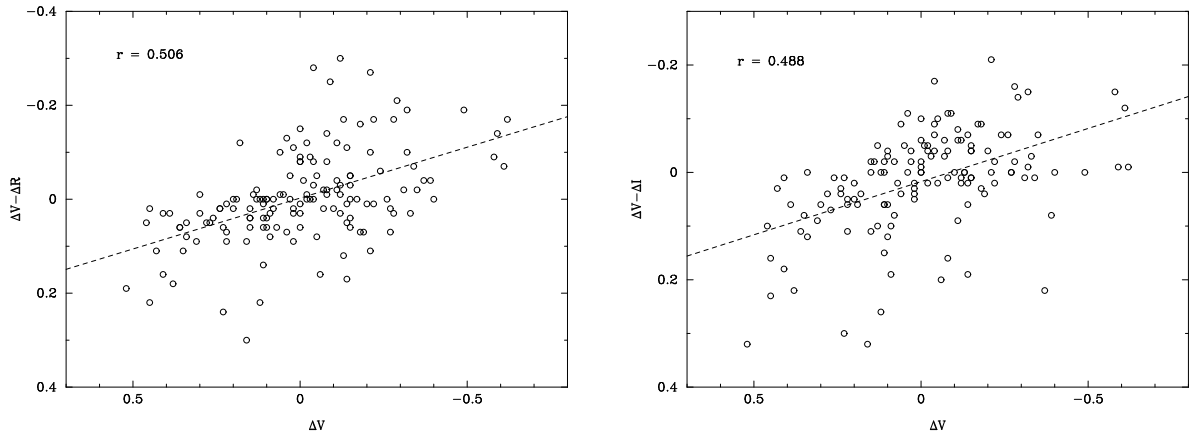}{f6b.ps}
\caption{The spectral behavior of the residual variations. The nightly-mean
residual magnitudes are used here. The dashed lines are the best fits to the
points and the correlation coefficients are respectively 0.506 and 0.488,
which indicate strong correlations or strong bluer-when-brighter chromatisms.}
\label{F6}
\end{figure}

\begin{figure}
\plotone{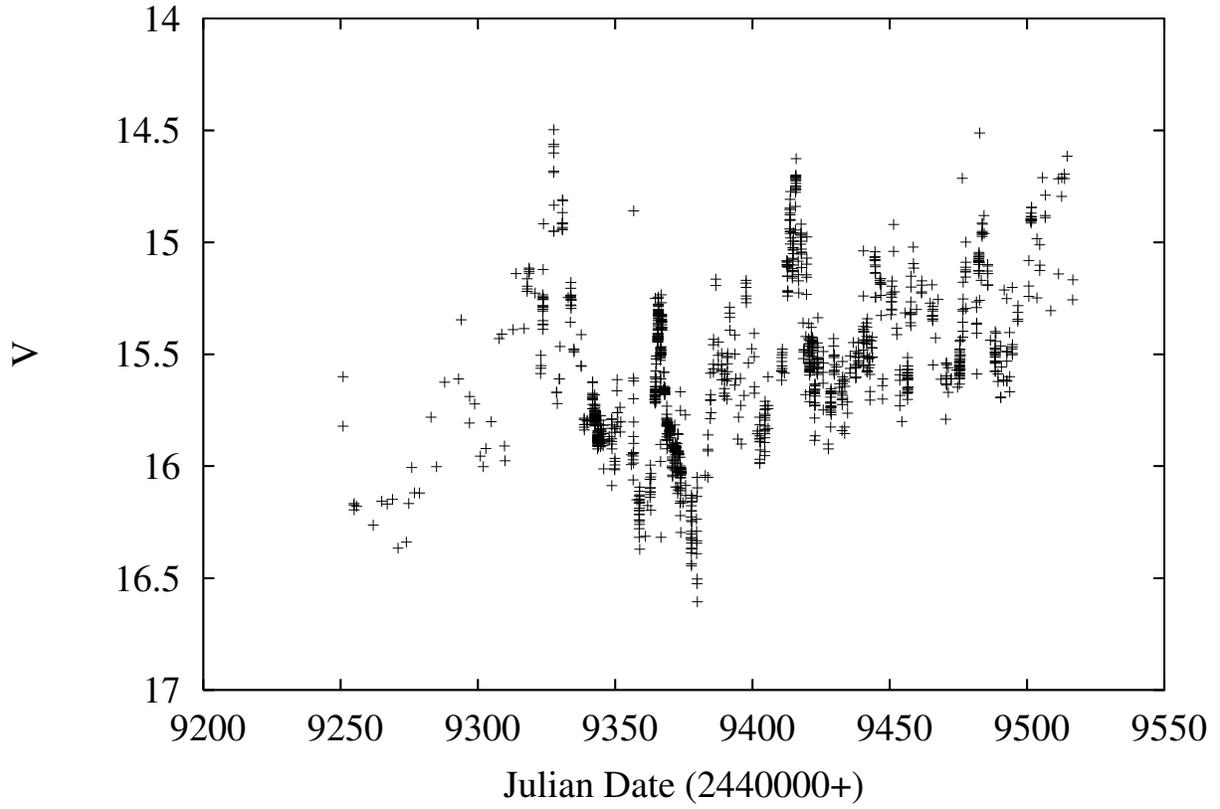}
\caption{The $V$-band light curve from 1993.7 to 1994.5 (data piece before
the first peak). It shows some signs of QPOs. See text for details. The
single point at (9356.5,14.86) is likely a false observation, so we do not
take it into consideration.}
\label{F7}
\end{figure}

That the 40-day period shows up in the variations in both quiescent and
outburst states gives us new insight into the prominent outburst of OJ~287.
It seems that the variation during the outburst phase can be resolved into
two components. The first component is the normal {\sl blazar} variation
(i.e., the residual variations obtained in \S4, see the smaller panels
in Fig.~\ref{F4}). It has the similar amplitude, period, and spectral
behavior as the variations in the quiescent state. The second component can
be taken as a `low-frequency modulation' (the solid lines in the larger
panels in Fig.~\ref{F4}) to the first component, and may be induced
by the interaction of the assumed BBHs in the center of this object
\citep{valtaoja00,liu02}.

The variability of blazars can be best explained with the shock-in-jet model
\citep{wagner95,marscher96}, although sometimes the geometric \citep[e.g.,][]
{wu05} or propagative \citep[e.g.][]{rickett01} effects, or some other
internal or external factors may also play a role. In the shock-in-jet model,
a twisted relativistic jet originates from the central black hole and contains
a hydromagnetically turbulent plasma. It undergoes fluctuations in its energy
input and this causes shock waves to develop and propagate down the jet.
Variability occurs when the shocks encounter fluctuations in the density of
relativistic electrons, in the magnitude of the magnetic field, and in the
orientation of the magnetic field.

For periodic variabilities, one usually turns to geometric origins, either
a precessing jet or light house effects \citep[e.g.,][]{camen92,katz97,
lainela99,wu05}. However, periodic variations resulting from the geometric
effects are likely to have a stable color. The bluer-when-brighter
chromatisms of OJ~287 reported in \S3.3 and \S4 suggest that both kinds of
periodic variations presented in this paper are not likely to resulted
from geometric effects. Also, the fact that the variations occur in the
optical regime and the presence of essentially simultaneous optical-radio
small flares \citep{pursimo00,valtaoja00} argue against a propagative origin
for them. 

In the first section, we have mentioned that \citet{marscher98} has proposed
that a ``duty cycle" of winding up of the magnetic field at the base of the
jet would result in major quasi-periodic injections of enhanced flow into the
jet \citep{ouyed97}. Here we will not discuss this possibility in accounting
for the 12-year period of OJ~287. But we suggest that Marscher's mechanism
does provide a good idea for explaining the 40-day's periodic variations of
this object. In fact, \citet{efimov02} have observed a 36.56-day's periodic
rotation of the plane of the polarization in OJ~287, which they considered as
a direct evidence for a helical magnetic field structure in the jet of this
object. The 36.56-day period is consistent with our roughly 40-day period,
and their observations provide a strong evidence for the scenario of winding
up of magnetic field at the base of the jet.

Another possibility for explaining the 40-day period may be related to the
orbital motion of the accretion disk around the central primary black hole of
OJ~287. At the redshift of 0.306, the 40-day period becomes $\sim30$ days at
the rest frame of OJ~287. Adopting a mass of $4\times10^8\,M_{\odot}$ for the
central primary black hole \citep{liu02}, the 30-day period corresponds to
the orbital motion at a radius of $r\sim17\,r_{\rm S}$, where $r_{\rm S}$ is
the Schwarzschild radius. This radius may represent the inner radius of the
accretion disk and the place from which the relativistic jet originates. (For
examples, according to recent numerical simulations, jets may originate from
several to 100 $r_{\rm S}$, see \citealt{meier01} and \citealt{hawley02}.)
Some disk oscillations at this radius may `propagate' and be kept in the jet,
and result in the observed 40-day period.

The 40-day periods both in the quiescent states and in the first peak in
late-1994 can be explained with the two scenarios described above. Then why
don't we observe the 40-day period in the second peak in 1995-96? According
to the models by \citet{valtaoja00} and \citet{liu02}, the structure of the
inner accretion disk is undisturbed in the quiescent state and in the first
peak of the major outbursts. However, when the effects of the impact of the
secondary black hole onto the primary accretion disk (see \S1) propagate to
the base of the primary jet ($\sim17\,r_{\rm S}$), the structure of the inner
primary accretion disk and the properties (electron density, magnetic field,
etc) of the primary jet are changed, and the accretion rate and hence the
jet emission are significantly enhanced. Then either the periodicity is
destroyed or the period changes to a new value. So we do not observe the
40-day period during the second peak of the major outburst.

After the second peak, the changed structure and properties of the primary
accretion disk and jet need some time to recover to the original states
\citep[in fact, the second peak is very broad and may extend to 1997, see
Fig.~2 in][]{pursimo00}, so the 40-day period was not observed during the
1--2 years after the second peak.

Our re-visitation of the OJ-94 data can put some constraints to the physical
model of OJ~287. In the old hit and penetration model by \citet{lehto96},
the precessing disk/jet model by \citet{katz97}, and the beaming model by
\citet{villata98}, the two peaks of the major outbursts are resulted from
the same physical processes and thus should have nearly the same behaviors.
Now the 40-day period shows up in the first peak but not in the second,
which gives further evidence for the assumption that the two peaks are
resulted from different physical processes. In other words, our results
are consistent with the models by \citet{valtaoja00} and \citet{liu02},
in which the first peak is thermal while the second is dominated by
synchrotron radiation. 

\section{SUMMARY}

During our monitoring of the BL Lac object OJ~287 in the first
half of 2005, the object did not show large-amplitude variations and was
in a relatively quiescent state. A possible period of 40 days was inferred
from its light curves. A bluer-when-brighter chromatism was found in the
variations, which is different from the overall spectral behavior during the
outburst state. The different spectral behaviors indicate different variation
mechanisms. The optical variability of OJ~287 during the OJ-94 project was
re-visited and again a probable 40-day period was
discovered. The physics responsible for the 40-day period is discussed. The
40-day period may be related to the helical structure of the magnetic field
at the base of the jet, or to the orbital motion close to the central primary
black hole.

\begin{figure}
\plotone{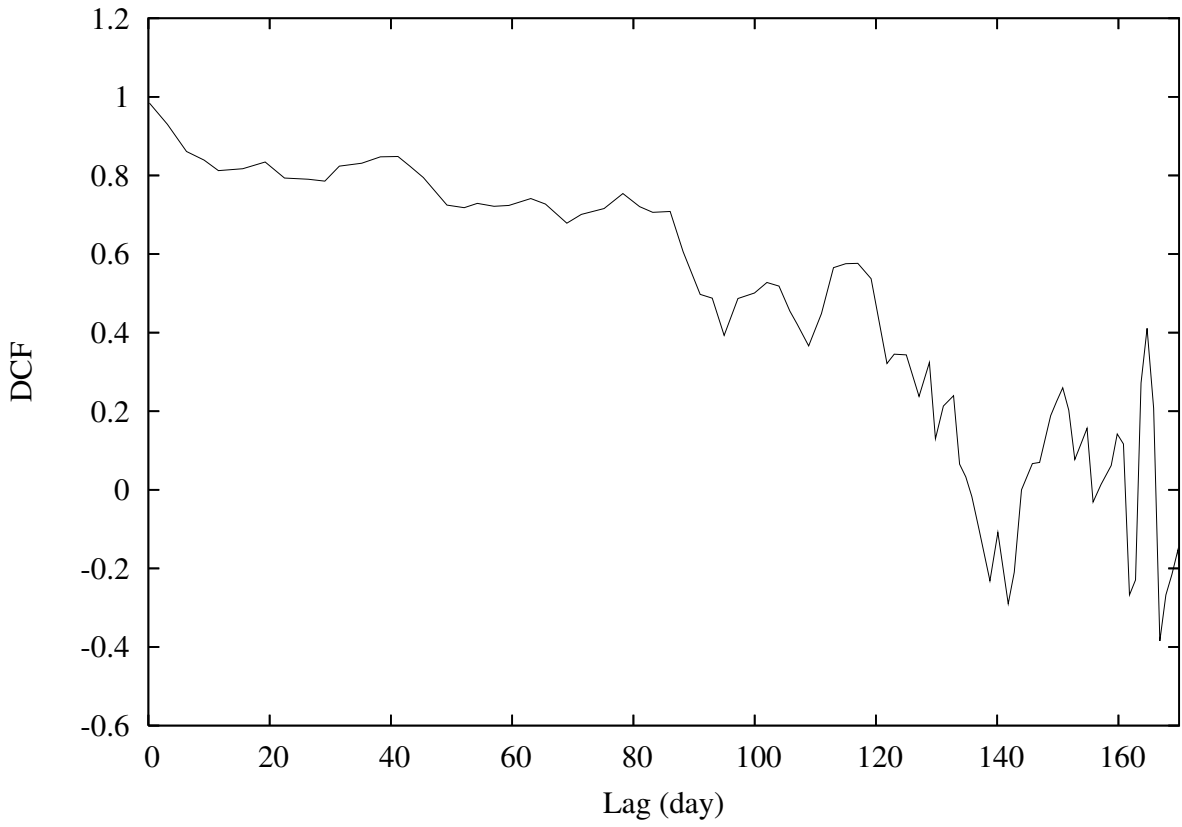}
\caption{ZDCF of the original light curve in the $R$ band in Fig.~\ref{F4}.
Compared to the right panels of Fig.~\ref{F5}, the maxima at 40, 80, and 120
days is not so evident, so one can hardly extract the 40-day period from
this figure.}
\label{F8}
\end{figure}

Except for the 36.56-day period discovered by \citet{efimov02}, the 40-day
period has not been reported before, even though OJ~287 has been observed for
more than 100 years. There are only very sparse photometric measurements
before the 1972 outburst (several or even only one measurement per year).
After that and till 1993, much denser monitorings were carried out but were
still not sufficient to reveal our claimed period of $\sim40$ days. The OJ-94
project, which
aimed at confirming the predicted outburst in 1994 and lasted from 1993 to
1997, provided the best opportunity to find the 40-day period. The reason
that no author has previously derived the 40-day period from the OJ-94 data
may be that a) the focus of the OJ-94 project is on the 12-year period, and
b) the methods to search for period in the variations (e.g., SF and ZDCF)
cannot give correct results when the prominent overall outburst with large
slope is not subtracted, as demonstrated by \citet{smith93} for the case of
SF and illustrated in Figure~\ref{F8} for the case of ZDCF.

Our monitoring revealed the 40-day period in the optical variability of
OJ~287 in its quiescent state, but the duration covers only two cycles and
there are two gaps in the light curves. The late-1994 outburst shows the
40-day period, but the major outbursts in 1971 and 1983 have sampling rates
too low to reveal this period. Although our monitoring results and the OJ-94
data do support each other, and suggest that the 40-day period is unlikely
to be a transient period, more data in both quiescent and outburst states
are needed to confirm this period. Fortunately, the
predicted next outburst of OJ~287 is in 2006 \citep{valtonen97,
kidger00, valtaoja00, liu02}. A large number of telescopes in the world will
surely monitor this object around the predicted time, and we will keep on
monitoring it intensively in order to confirm the 40-day period both in the
quiescent and outburst states.

\begin{acknowledgements}
The authors thank the anonymous referee for insightful comments and
suggestions that helped to improve this paper very much. We are in debt to
L. I. Gurvits who has read part of this paper and given
some valuable comments. This work has made use of the data from the OJ-94
project (http://www.astro.utu.fi/oj94/), and has been supported by the
Chinese National Natural Science Foundation, No. 10473012, 10573020,
10303003, 10473001, 10525313, and 10203001.
\end{acknowledgements}

\clearpage

\clearpage

\begin{deluxetable}{cccccc}
\tablewidth{0pt}
\tablecaption{Observational log and results in the BATC $e$ band.\tablenotemark{a} \label{T1}}
\tablehead{\colhead{Obs. Date\tablenotemark{b}} & \colhead{Obs. Time\tablenotemark{b}} & \colhead{Julian Date} & \colhead{Exp. Time} & \colhead{$e$} & \colhead{$e_{\rm err}$} \\
\colhead{(yyyy mm dd)} & \colhead{(hh:mm:ss)} &  & \colhead{(s)} & \colhead{(mag)} & \colhead{(mag)}
}
\startdata
 2005 01 29 & 14:57:32.0 & 2453400.123 & 300 & 15.671 & 0.144 \\
 2005 01 29 & 15:24:21.0 & 2453400.142 & 480 & 15.425 & 0.119 \\
 2005 01 30 & 16:32:48.0 & 2453401.189 & 240 & 15.602 & 0.064 \\
 2005 01 30 & 16:44:07.0 & 2453401.197 & 240 & 15.519 & 0.060 \\
 2005 01 30 & 16:58:03.0 & 2453401.207 & 240 & 15.525 & 0.057 \\
 2005 01 30 & 17:11:56.0 & 2453401.217 & 240 & 15.539 & 0.070 \\
 2005 01 30 & 17:26:02.0 & 2453401.226 & 240 & 15.394 & 0.058 \\
 2005 01 30 & 17:40:07.0 & 2453401.236 & 240 & 15.451 & 0.058 \\
 2005 01 30 & 17:53:57.0 & 2453401.246 & 240 & 15.497 & 0.061 \\
 2005 01 30 & 18:07:58.0 & 2453401.256 & 240 & 15.610 & 0.064 \\
 2005 01 30 & 18:21:47.0 & 2453401.265 & 240 & 15.538 & 0.062 \\
 2005 01 30 & 18:35:42.0 & 2453401.275 & 240 & 15.362 & 0.051 \\
 2005 01 30 & 18:49:24.0 & 2453401.284 & 240 & 15.431 & 0.055 \\
 2005 02 01 & 18:10:46.0 & 2453403.258 & 240 & 15.859 & 0.081 \\
 2005 02 02 & 13:55:03.0 & 2453404.080 & 240 & 15.631 & 0.069 \\
 2005 02 18 & 14:44:37.0 & 2453420.114 & 240 & 15.256 & 0.063 \\
 2005 02 19 & 12:48:06.0 & 2453421.033 & 240 & 15.066 & 0.058 \\
 2005 02 24 & 12:02:17.0 & 2453426.001 & 240 & 14.954 & 0.083 \\
 2005 02 24 & 12:16:13.0 & 2453426.011 & 240 & 14.797 & 0.068 \\
 2005 02 24 & 12:30:16.0 & 2453426.021 & 240 & 14.794 & 0.062 \\
 2005 02 24 & 12:44:14.0 & 2453426.031 & 240 & 14.926 & 0.074 \\
 2005 02 26 & 12:55:37.0 & 2453428.039 & 240 & 15.092 & 0.070 \\
 2005 02 27 & 13:39:46.0 & 2453429.069 & 240 & 15.168 & 0.056 \\
 2005 03 05 & 11:53:22.0 & 2453434.995 & 240 & 15.040 & 0.055 \\
 2005 03 06 & 14:17:23.0 & 2453436.095 & 300 & 14.697 & 0.124 \\
 2005 03 06 & 17:33:11.0 & 2453436.231 & 300 & 15.086 & 0.043 \\
 2005 03 06 & 17:51:14.0 & 2453436.244 & 300 & 15.066 & 0.041 \\
 2005 03 14 & 13:54:54.0 & 2453444.080 & 240 & 15.256 & 0.184 \\
 2005 03 15 & 14:19:50.0 & 2453445.097 & 240 & 15.409 & 0.076 \\
 2005 03 18 & 13:15:05.0 & 2453448.052 & 240 & 15.857 & 0.141 \\
 2005 03 18 & 13:35:35.0 & 2453448.066 & 240 & 15.981 & 0.131 \\
 2005 03 18 & 14:51:41.0 & 2453448.119 & 240 & 15.698 & 0.084 \\
 2005 03 18 & 15:05:46.0 & 2453448.129 & 240 & 15.542 & 0.083 \\
 2005 03 18 & 15:19:38.0 & 2453448.139 & 240 & 15.472 & 0.076 \\
 2005 03 18 & 15:32:58.0 & 2453448.148 & 240 & 15.527 & 0.079 \\
 2005 03 19 & 14:00:39.0 & 2453449.084 & 240 & 15.554 & 0.133 \\
 2005 03 19 & 14:14:30.0 & 2453449.094 & 240 & 15.620 & 0.158 \\
 2005 03 19 & 14:28:37.0 & 2453449.103 & 240 & 15.775 & 0.166 \\
 2005 03 19 & 14:42:45.0 & 2453449.113 & 240 & 16.115 & 0.245 \\
 2005 03 23 & 11:56:22.0 & 2453452.998 & 240 & 15.277 & 0.110 \\
 2005 03 26 & 14:08:02.0 & 2453456.089 & 240 & 15.007 & 0.097 \\
 2005 03 27 & 14:18:35.0 & 2453457.096 & 240 & 15.067 & 0.137 \\
 2005 03 30 & 11:26:04.0 & 2453459.976 & 240 & 14.923 & 0.076 \\
 2005 03 30 & 11:40:02.0 & 2453459.986 & 240 & 14.724 & 0.047 \\
 2005 03 30 & 11:54:03.0 & 2453459.996 & 240 & 14.822 & 0.049 \\
 2005 03 30 & 12:08:01.0 & 2453460.006 & 240 & 14.841 & 0.045 \\
 2005 03 30 & 12:22:07.0 & 2453460.015 & 240 & 14.962 & 0.054 \\
 2005 03 30 & 12:36:10.0 & 2453460.025 & 240 & 15.014 & 0.049 \\
 2005 03 30 & 12:50:05.0 & 2453460.035 & 240 & 14.924 & 0.053 \\
 2005 03 30 & 13:04:11.0 & 2453460.045 & 240 & 14.810 & 0.054 \\
 2005 04 21 & 12:55:57.0 & 2453482.039 & 240 & 15.186 & 0.084 \\
 2005 04 21 & 13:10:03.0 & 2453482.049 & 240 & 15.183 & 0.082 \\
 2005 04 22 & 13:24:51.0 & 2453483.059 & 240 & 15.547 & 0.125 \\
 2005 04 22 & 13:38:52.0 & 2453483.069 & 240 & 15.295 & 0.105 \\
 2005 04 22 & 13:52:31.0 & 2453483.078 & 240 & 15.717 & 0.179 \\
 2005 04 24 & 11:59:37.0 & 2453485.000 & 240 & 15.484 & 0.100 \\
 2005 04 24 & 12:13:39.0 & 2453485.010 & 240 & 15.822 & 0.123 \\
 2005 04 24 & 12:27:37.0 & 2453485.019 & 240 & 15.587 & 0.100 \\
 2005 04 24 & 12:41:45.0 & 2453485.029 & 240 & 15.609 & 0.098 \\
 2005 04 24 & 12:55:48.0 & 2453485.039 & 240 & 15.523 & 0.102 \\
 2005 04 28 & 12:47:45.0 & 2453489.033 & 240 & 16.140 & 0.144 \\
\enddata
\tablenotetext{a}{We took part in the multiwavelength campaign on OJ~287
coordinated by Stefano Ciprini during 2005 April 13--15 (or JD~2,453,473--475),
so we do not present the data in these days in Tables~1--3 but give a point
(the mean) per night in each light curve in Fig.~\ref{F1}.}
\tablenotetext{b}{The obs. date and time are of universal time. The same is
for Tables 2 and 3.}
\end{deluxetable}

\clearpage

\begin{deluxetable}{cccccc}
\tablewidth{0pt}
\tabletypesize{\small}
\tablecaption{Observational log and results in the BATC $i$ band.\label{T2}}
\tablehead{\colhead{Obs. Date} & \colhead{Obs. Time} & \colhead{Julian Date} & \colhead{Exp. Time} & \colhead{$i$} & \colhead{$i_{\rm err}$} \\
\colhead{(yyyy mm dd)} & \colhead{(hh:mm:ss)} &  & \colhead{(s)} & \colhead{(mag)} & \colhead{(mag)}
}
\startdata
 2005 01 29 & 15:02:48.0 & 2453400.127 & 180 & 14.637 & 0.048 \\
 2005 01 30 & 16:28:41.0 & 2453401.187 & 150 & 14.646 & 0.023 \\
 2005 01 30 & 16:48:34.0 & 2453401.200 & 150 & 14.670 & 0.026 \\
 2005 01 30 & 17:02:23.0 & 2453401.210 & 150 & 14.168 & 0.033 \\
 2005 01 30 & 17:16:22.0 & 2453401.220 & 150 & 14.661 & 0.024 \\
 2005 01 30 & 17:30:36.0 & 2453401.229 & 150 & 14.620 & 0.026 \\
 2005 01 30 & 17:44:21.0 & 2453401.239 & 150 & 14.676 & 0.026 \\
 2005 01 30 & 17:58:24.0 & 2453401.249 & 150 & 14.710 & 0.027 \\
 2005 01 30 & 18:12:23.0 & 2453401.259 & 150 & 14.697 & 0.026 \\
 2005 01 30 & 18:26:11.0 & 2453401.268 & 150 & 14.607 & 0.022 \\
 2005 01 30 & 18:39:56.0 & 2453401.278 & 150 & 14.602 & 0.023 \\
 2005 01 30 & 18:53:50.0 & 2453401.287 & 150 & 14.681 & 0.028 \\
 2005 01 31 & 16:56:13.0 & 2453402.206 & 150 & 15.070 & 0.058 \\
 2005 01 31 & 17:54:51.0 & 2453402.246 & 240 & 14.948 & 0.034 \\
 2005 02 01 & 18:06:20.0 & 2453403.254 & 150 & 15.128 & 0.032 \\
 2005 02 02 & 13:50:34.0 & 2453404.077 & 150 & 14.856 & 0.027 \\
 2005 02 18 & 14:33:24.0 & 2453420.106 & 150 & 14.454 & 0.026 \\
 2005 02 19 & 12:55:55.0 & 2453421.039 & 150 & 14.321 & 0.031 \\
 2005 02 24 & 12:06:44.0 & 2453426.005 & 150 & 14.227 & 0.040 \\
 2005 02 24 & 12:20:41.0 & 2453426.014 & 150 & 14.117 & 0.038 \\
 2005 02 24 & 12:34:42.0 & 2453426.024 & 150 & 14.200 & 0.043 \\
 2005 02 24 & 12:48:38.0 & 2453426.034 & 150 & 14.203 & 0.036 \\
 2005 02 26 & 13:00:04.0 & 2453428.042 & 150 & 14.297 & 0.032 \\
 2005 02 27 & 13:44:13.0 & 2453429.072 & 150 & 14.308 & 0.021 \\
 2005 03 05 & 11:43:47.0 & 2453434.989 & 150 & 14.514 & 0.040 \\
 2005 03 06 & 14:06:29.0 & 2453436.088 & 150 & 14.457 & 0.089 \\
 2005 03 14 & 13:59:20.0 & 2453444.083 & 150 & 14.642 & 0.117 \\
 2005 03 15 & 14:24:16.0 & 2453445.100 & 150 & 14.726 & 0.032 \\
 2005 03 18 & 13:19:33.0 & 2453448.055 & 150 & 15.028 & 0.131 \\
 2005 03 18 & 13:40:04.0 & 2453448.070 & 150 & 14.571 & 0.112 \\
 2005 03 18 & 14:56:06.0 & 2453448.122 & 150 & 14.922 & 0.043 \\
 2005 03 18 & 15:10:14.0 & 2453448.132 & 150 & 14.968 & 0.049 \\
 2005 03 18 & 15:23:49.0 & 2453448.142 & 150 & 14.893 & 0.046 \\
 2005 03 18 & 15:37:23.0 & 2453448.151 & 150 & 14.823 & 0.038 \\
 2005 03 19 & 14:04:53.0 & 2453449.087 & 150 & 14.965 & 0.080 \\
 2005 03 19 & 14:18:57.0 & 2453449.096 & 150 & 14.898 & 0.075 \\
 2005 03 19 & 14:33:05.0 & 2453449.106 & 150 & 14.935 & 0.075 \\
 2005 03 19 & 14:47:12.0 & 2453449.116 & 150 & 14.995 & 0.092 \\
 2005 03 23 & 12:00:49.0 & 2453453.000 & 150 & 14.703 & 0.063 \\
 2005 03 26 & 14:12:41.0 & 2453456.092 & 150 & 14.313 & 0.047 \\
 2005 03 27 & 14:13:52.0 & 2453457.093 & 150 & 14.218 & 0.054 \\
 2005 03 30 & 11:30:25.0 & 2453459.979 & 150 & 14.298 & 0.024 \\
 2005 03 30 & 11:44:28.0 & 2453459.989 & 150 & 14.195 & 0.024 \\
 2005 03 30 & 11:58:28.0 & 2453459.999 & 150 & 14.214 & 0.023 \\
 2005 03 30 & 12:12:31.0 & 2453460.009 & 150 & 14.208 & 0.023 \\
 2005 03 30 & 12:26:35.0 & 2453460.019 & 150 & 14.212 & 0.024 \\
 2005 03 30 & 12:40:32.0 & 2453460.028 & 150 & 14.219 & 0.021 \\
 2005 03 30 & 12:54:30.0 & 2453460.038 & 150 & 14.160 & 0.020 \\
 2005 03 30 & 13:08:34.0 & 2453460.048 & 150 & 14.151 & 0.022 \\
 2005 04 21 & 13:00:28.0 & 2453482.042 & 150 & 14.472 & 0.038 \\
 2005 04 21 & 13:14:26.0 & 2453482.052 & 150 & 14.475 & 0.043 \\
 2005 04 22 & 13:29:17.0 & 2453483.062 & 150 & 14.541 & 0.045 \\
 2005 04 22 & 13:43:17.0 & 2453483.072 & 150 & 14.617 & 0.047 \\
 2005 04 22 & 13:56:56.0 & 2453483.081 & 150 & 14.644 & 0.046 \\
 2005 04 24 & 12:04:30.0 & 2453485.003 & 150 & 14.864 & 0.044 \\
 2005 04 24 & 12:17:54.0 & 2453485.012 & 150 & 14.925 & 0.051 \\
 2005 04 24 & 12:32:06.0 & 2453485.022 & 150 & 15.052 & 0.052 \\
 2005 04 24 & 12:46:16.0 & 2453485.032 & 150 & 14.958 & 0.054 \\
 2005 04 24 & 13:00:12.0 & 2453485.042 & 150 & 14.891 & 0.044 \\
 2005 04 28 & 12:52:12.0 & 2453489.036 & 150 & 15.505 & 0.067 \\
\enddata
\end{deluxetable}

\clearpage

\begin{deluxetable}{cccccc}
\tabletypesize{\small}
\tablewidth{0pt}
\tablecaption{Observational log and results in the BATC $m$ band.\label{T3}}
\tablehead{\colhead{Obs. Date} & \colhead{Obs. Time} & \colhead{Julian Date} & \colhead{Exp. Time} & \colhead{$m$} & \colhead{$m_{\rm err}$} \\
\colhead{(yyyy mm dd)} & \colhead{(hh:mm:ss)} &  & \colhead{(s)} & \colhead{(mag)} & \colhead{(mag)}
}
\startdata
 2005 01 30 & 16:38:03.0 & 2453401.193 & 240 & 14.360 & 0.027 \\
 2005 01 30 & 16:52:59.0 & 2453401.203 & 240 & 14.259 & 0.028 \\
 2005 01 30 & 17:06:48.0 & 2453401.213 & 240 & 14.224 & 0.026 \\
 2005 01 30 & 17:20:50.0 & 2453401.223 & 240 & 14.280 & 0.025 \\
 2005 01 30 & 17:35:00.0 & 2453401.233 & 240 & 14.255 & 0.030 \\
 2005 01 30 & 17:48:47.0 & 2453401.242 & 240 & 14.330 & 0.029 \\
 2005 01 30 & 18:02:51.0 & 2453401.252 & 240 & 14.288 & 0.027 \\
 2005 01 30 & 18:16:35.0 & 2453401.261 & 240 & 14.333 & 0.029 \\
 2005 01 30 & 18:30:30.0 & 2453401.271 & 240 & 14.256 & 0.026 \\
 2005 01 30 & 18:44:22.0 & 2453401.281 & 240 & 14.271 & 0.028 \\
 2005 01 30 & 18:58:11.0 & 2453401.290 & 240 & 14.242 & 0.026 \\
 2005 01 31 & 18:07:54.0 & 2453402.255 & 300 & 14.411 & 0.043 \\
 2005 02 01 & 17:54:36.0 & 2453403.246 & 240 & 14.632 & 0.031 \\
 2005 02 02 & 13:45:46.0 & 2453404.073 & 240 & 14.542 & 0.035 \\
 2005 02 18 & 14:38:37.0 & 2453420.110 & 240 & 14.168 & 0.028 \\
 2005 02 19 & 13:05:40.0 & 2453421.046 & 240 & 13.938 & 0.029 \\
 2005 02 24 & 12:11:08.0 & 2453426.008 & 240 & 13.911 & 0.039 \\
 2005 02 24 & 12:25:04.0 & 2453426.017 & 240 & 13.866 & 0.044 \\
 2005 02 24 & 12:39:06.0 & 2453426.027 & 240 & 13.775 & 0.035 \\
 2005 02 24 & 12:53:01.0 & 2453426.037 & 240 & 13.887 & 0.038 \\
 2005 02 26 & 13:04:16.0 & 2453428.045 & 240 & 14.008 & 0.032 \\
 2005 02 27 & 13:48:40.0 & 2453429.075 & 240 & 13.979 & 0.023 \\
 2005 03 05 & 11:48:12.0 & 2453434.992 & 240 & 14.129 & 0.041 \\
 2005 03 06 & 14:10:55.0 & 2453436.091 & 240 & 14.171 & 0.101 \\
 2005 03 15 & 14:28:38.0 & 2453445.103 & 240 & 14.346 & 0.034 \\
 2005 03 18 & 13:30:25.0 & 2453448.063 & 240 & 14.696 & 0.092 \\
 2005 03 18 & 13:46:21.0 & 2453448.074 & 240 & 14.618 & 0.209 \\
 2005 03 18 & 14:46:31.0 & 2453448.116 & 240 & 14.605 & 0.040 \\
 2005 03 18 & 15:00:33.0 & 2453448.125 & 240 & 14.627 & 0.042 \\
 2005 03 18 & 15:14:41.0 & 2453448.135 & 240 & 14.493 & 0.050 \\
 2005 03 18 & 15:27:58.0 & 2453448.145 & 240 & 14.593 & 0.042 \\
 2005 03 19 & 14:09:20.0 & 2453449.090 & 240 & 14.704 & 0.083 \\
 2005 03 19 & 14:23:27.0 & 2453449.100 & 240 & 14.569 & 0.077 \\
 2005 03 19 & 14:37:32.0 & 2453449.109 & 240 & 14.675 & 0.075 \\
 2005 03 19 & 14:51:27.0 & 2453449.119 & 240 & 14.449 & 0.071 \\
 2005 03 23 & 12:10:44.0 & 2453453.008 & 240 & 14.420 & 0.063 \\
 2005 03 26 & 14:17:09.0 & 2453456.095 & 240 & 14.247 & 0.052 \\
 2005 03 27 & 14:08:19.0 & 2453457.089 & 240 & 13.949 & 0.046 \\
 2005 03 30 & 11:34:53.0 & 2453459.983 & 240 & 13.908 & 0.027 \\
 2005 03 30 & 11:48:55.0 & 2453459.992 & 240 & 13.910 & 0.025 \\
 2005 03 30 & 12:02:52.0 & 2453460.002 & 240 & 13.859 & 0.025 \\
 2005 03 30 & 12:16:56.0 & 2453460.012 & 240 & 13.886 & 0.023 \\
 2005 03 30 & 12:31:01.0 & 2453460.021 & 240 & 13.867 & 0.024 \\
 2005 03 30 & 12:44:57.0 & 2453460.031 & 240 & 13.930 & 0.026 \\
 2005 03 30 & 12:58:59.0 & 2453460.041 & 240 & 13.857 & 0.024 \\
 2005 03 30 & 13:13:01.0 & 2453460.051 & 240 & 13.882 & 0.026 \\
 2005 04 21 & 13:04:55.0 & 2453482.045 & 240 & 14.169 & 0.043 \\
 2005 04 21 & 13:18:51.0 & 2453482.055 & 240 & 14.085 & 0.045 \\
 2005 04 22 & 13:33:42.0 & 2453483.065 & 240 & 14.326 & 0.051 \\
 2005 04 22 & 13:47:34.0 & 2453483.075 & 240 & 14.173 & 0.045 \\
 2005 04 22 & 14:01:18.0 & 2453483.084 & 240 & 14.372 & 0.049 \\
 2005 04 24 & 12:08:38.0 & 2453485.006 & 240 & 14.694 & 0.052 \\
 2005 04 24 & 12:22:22.0 & 2453485.016 & 240 & 14.602 & 0.054 \\
 2005 04 24 & 12:36:33.0 & 2453485.025 & 240 & 14.429 & 0.043 \\
 2005 04 24 & 12:50:42.0 & 2453485.035 & 240 & 14.544 & 0.050 \\
 2005 04 24 & 13:04:26.0 & 2453485.045 & 240 & 14.581 & 0.052 \\
 2005 04 28 & 12:56:35.0 & 2453489.039 & 240 & 15.025 & 0.083 \\
\enddata
\end{deluxetable}

\clearpage

\begin{deluxetable}{lccc}
\tablewidth{0pt}
\tablecaption{BATC magnitudes of the three comparison stars\label{T4}}
\tablehead{\colhead{Star ID} & \colhead{$e$} & \colhead{$i$} & \colhead{$m$} \\
 & \colhead{(mag)} & \colhead{(mag)} & \colhead{(mag)} }
\startdata
 4 & 14.920 & 14.020 & 13.650 \\
10 & 15.010 & 14.510 & 14.280 \\
11 & 15.490 & 14.840 & 14.590 \\
\enddata
\end{deluxetable}

\end{document}